\documentclass[prb,aps, twocolumn, amsmath,amssymb,superscriptaddress]{revtex4}
\usepackage{graphicx} % Required for inserting images
\usepackage{float}
\usepackage{amsmath}
\usepackage{amssymb}
\usepackage{amsfonts}
\usepackage{euscript}
\usepackage{enumerate}
\usepackage{hhline}
\usepackage{pslatex}
\usepackage{tabularx}
\usepackage[usenames,dvipsnames]{xcolor}

\usepackage[colorlinks,linkcolor=red,anchorcolor=blue,citecolor=blue]{hyperref}

\makeatletter
\renewcommand{\@biblabel}[1]{#1. }
\renewcommand{\@dotsep}{500}
\renewcommand{\@pnumwidth}{0em}
\renewcommand{\l@figure}[2]{% #1 is e.g. Figure 1 + caption, #2 is pg.
\@dottedtocline{1}{1.5em}{2em}{Figure #1}{}\vspace{15pt}}

\begin{document}

\title{Super-resolved snapshot hyperspectral imaging of solid-state quantum emitters for high-throughput integrated quantum technologies}

\author{Shunfa Liu}
\affiliation{State Key Laboratory of Optoelectronic Materials and Technologies, School of Physics, Sun Yat-sen University, Guangzhou 510275, China}

\author{Xueshi Li}
\affiliation{State Key Laboratory of Optoelectronic Materials and Technologies, School of Physics, Sun Yat-sen University, Guangzhou 510275, China}

\author{Hanqing Liu}
\affiliation{State Key Laboratory for Superlattice and Microstructures, Institute of Semiconductors, Chinese Academy of Sciences, Beijing 100083, China.}
\affiliation{Center of Materials Science and Optoelectronics Engineering, University of Chinese Academy of Sciences, Beijing 100049, China.}

\author{Guixin Qiu}
\affiliation{State Key Laboratory of Optoelectronic Materials and Technologies, School of Physics, Sun Yat-sen University, Guangzhou 510275, China}

\author{Jiantao Ma}
\affiliation{State Key Laboratory of Optoelectronic Materials and Technologies, School of Physics, Sun Yat-sen University, Guangzhou 510275, China}

\author{Liang Nie}
\affiliation{State Key Laboratory of Optoelectronic Materials and Technologies, School of Physics, Sun Yat-sen University, Guangzhou 510275, China}

\author{\\Haiqiao Ni}
\affiliation{State Key Laboratory for Superlattice and Microstructures, Institute of Semiconductors, Chinese Academy of Sciences, Beijing 100083, China.}
\affiliation{Center of Materials Science and Optoelectronics Engineering, University of Chinese Academy of Sciences, Beijing 100049, China.}

\author{Zhichuan Niu}
\affiliation{State Key Laboratory for Superlattice and Microstructures, Institute of Semiconductors, Chinese Academy of Sciences, Beijing 100083, China.}
\affiliation{Center of Materials Science and Optoelectronics Engineering, University of Chinese Academy of Sciences, Beijing 100049, China.}

\author{Cheng-Wei Qiu}
\affiliation{Department of Electrical and Computer Engineering, National University of Singapore, Singapore, Singapore.}

\author{Xuehua Wang}
\thanks{wangxueh@mail.sysu.edu.cn}
\affiliation{State Key Laboratory of Optoelectronic Materials and Technologies, School of Physics, Sun Yat-sen University, Guangzhou 510275, China}

\author{Jin Liu}
\thanks{liujin23@mail.sysu.edu.cn}
\affiliation{State Key Laboratory of Optoelectronic Materials and Technologies, School of Physics, Sun Yat-sen University, Guangzhou 510275, China}
\date{\today}

\begin{abstract}
\noindent \textbf{Solid-state quantum emitters coupled to integrated photonic nanostructures are quintessential for exploring fundamental phenomena in cavity quantum electrodynamics and widely employed in photonic quantum technologies such as non-classical light sources, quantum repeaters, and quantum transducers, etc. One of the most exciting promises from integrated quantum photonics is the potential of scalability that enables massive productions of miniaturized devices on a single chip. In reality, the yield of efficient and reproducible light-matter couplings is greatly hindered by the spectral and spatial mismatches between the single solid-state quantum emitters and confined or propagating optical modes supported by the photonic nanostructures, preventing the high-throughput realization of large-scale integrated quantum photonic circuits for more advanced quantum information processing tasks. In this work, we introduce the concept of hyperspectral imaging in quantum optics, for the first time, to address such a long-standing issue. By exploiting the extended mode with a unique dispersion in a 1D planar cavity, the spectral and spatial information of each individual quantum dot in an ensemble can be accurately and reliably extracted from a single wide-field photoluminescence image with super-resolutions. With the extracted quantum dot positions and emission wavelengths, surface-emitting quantum light sources and in-plane photonic circuits can be deterministically fabricated with a high-throughput by etching the 1D confined planar cavity into 3D confined micropillars and 2D confined waveguides. Further extension of this technique by employing an open planar cavity could be exploited for pursuing a variety of compact quantum photonic devices with expanded functionalities for large-scale integration. Our work is expected to change the landscape of integrated quantum photonic technology in which solid-state quantum emitters play essential roles as superior quantum light sources and efficient spin-photon interfaces.}
\end{abstract}

\maketitle

Solid-state quantum emitters play an essential role in exploring quantum physics~\cite{schulte2015quadrature,tiranov2023collective,tomm2023photon} and developing integrated photonic quantum technologies~\cite{o2009photonic,wang2020integrated,uppu2021quantum,zhou2023epitaxial}. The efficient coupling of solid-state quantum emitters to on-chip photonic nanostructures offers the ability to understand and harness the nanoscale interactions between light and matter at the single-photon level~\cite{englund2007controlling,lodahl2015interfacing,javadi2015single,sun2018single,Li2023}. In particular, coupling epitaxial quantum dots (QDs) to a micro-resonator facilitates the realization of near-optimal quantum light sources with simultaneous degrees of source brightness, single-photon purity, entanglement fidelity, and photon indistinguishability~\cite{somaschi2016near,he2017deterministic,wang2019towards,liu2019solid,wang2019demand,uppu2020scalable,tomm2021bright}. Such unprecedented sources of non-classic light exhibit unique advantages in the implementations of quantum communication~\cite{vajner2022quantum}, quantum simulation~\cite{wang2019boson}, and quantum metrology~\cite{muller2017quantum,wang2020observation}. Moving towards a quantum network, spin-photon entanglements~\cite{gao2012observation,de2012quantum} in coupled QD-cavity systems have been exploited for building quantum repeaters~\cite{azuma2022quantum} based on either quantum memories~\cite{afzelius2015quantum} or photonic cluster states~\cite{coste2023high,cogan2023deterministic}. The aforementioned achievements heavily rely on deterministic light-matter couplings which require scalable realizations of both spatial and spectral overlaps between the quantum emitters and the optical modes supported by the photonic nanostructures~\cite{liu2021nanoscale}. 
\begin{center}
	\begin{figure*}
		\begin{center}
			\includegraphics[width=\linewidth]{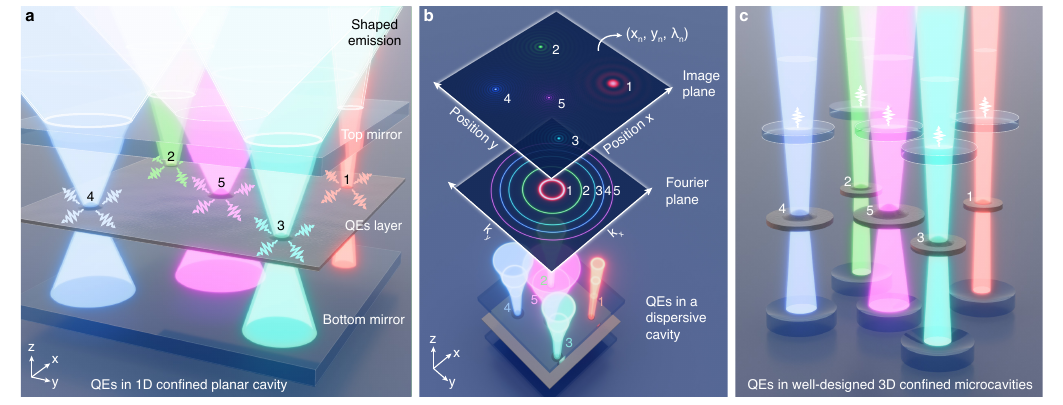}
			\caption{\textbf{Emission characteristics of single solid-state quantum emitters in a 1D nanophotonic planar cavity and a 3D confined micro-cavity.} (a) A multitude of single solid-state quantum emitters embedded in a 1D nanophotonic planar cavity supporting an optical mode confined in the Z direction while extended in the XY plane. (b) Far-field patterns and image profiles of different solid-state quantum emitters coupled to the dispersive nanophotonic planar cavity. The spatial position and emission wavelength ($\rm{x_{n}},\rm{y_{n}},\lambda_{n}$) of each individual quantum emitter can be extracted from the wide-field image. (c) Deterministic couplings between QDs and 3D confined microcavities, showing emissions with enhanced radiative rates and improved directionalities.}
			\label{fig:Fig1}
		\end{center}
	\end{figure*}
\end{center}
Taking a cavity quantum electrodynamics (QED) system, consisting of a charged exciton in a QD coupled to a micro-resonator, as an example, the figure-of-merit for light-matter coupling is the so-called Purcell factor~\cite{gerard1999strong} $F_{p}= \frac{3(\lambda_c/n)^3}{4\pi^2}\times \frac{Q}{V}\times \frac{\Delta \lambda_c^2}{4(\lambda_0-\lambda_c)^2+\Delta\lambda_c^2} \times \frac{|\vec{E}(\vec{r})|^2}{|\vec{E}_{\rm{max}}|^2}$, in which $\lambda_c$ and n are the cavity resonant wavelength and refractive index of the cavity material, Q and V are the Q-factor and mode volume of the cavity, $\lambda_0$ is the wavelength of the emitter, $\Delta\lambda_c$ is the full-width half maximum (FWHM) of the cavity mode, $\vec{E}(\vec{r})$ and $\vec{E}_{\rm{max}}$ are the local electric field at the QD position and the cavity mode anti-node, respectively. In order to reach the maximal Purcell factor provided by a high-Q and low mode-volume cavity, one needs to accurately place the QD into the cavity field maxima ($\vec{E}(\vec{r})=\vec{E}_{\rm{max}}$) and precisely align the QD emission wavelength to the cavity resonance ($\lambda_0=\lambda_c$). However, the naturally efficient light-matter couplings are intrinsic low-probability events due to the fact that solid-state quantum emitters universally suffer from both spatial and spectral inhomogeneities. Specifically, epitaxial InAs QDs are randomly distributed within the plane of the host material and exhibit an appreciable spectral inhomogeneous broadening which is typically orders of magnitude larger than the homogeneous linewidth of single QDs~\cite{englund2005controlling}. Through continuous technological developments in the past decade, the spatial matching condition is now steadily accessible by using atomic force microscopy\cite{badolato2005deterministic,hennessy2007quantum}, scanning electron microscopy\cite{kuruma2016position}, confocal optical microscopy\cite{thon2009strong}, in-situ optical lithography\cite{dousse2008controlled,kolatschek2019deterministic}, in-situ electron beam lithography\cite{gschrey2015highly}, and photoluminescence (PL) imaging\cite{kojima2013accurate,sapienza2015nanoscale,liu2017cryogenic,pregnolato2020deterministic}. In particular, wide-field PL imaging exhibits the advantages of parallelly extracting the spatial positions of a large number of quantum emitters with an accuracy of deep sub-wavelength by using the super-localization technique\cite{yildiz2003myosin}. However, the bottleneck of high-throughput deterministic couplings is the achievement of spectral alignments between the emission wavelengths of quantum emitters and the operation bands of nanophotonic structures. For achieving the spectral matching condition, the spectrum of individual quantum emitter has to be firstly measured in a one-by-one scanning manner by using a raster scanning confocal $\mu$PL/$\mu$EL spectroscopy, which makes the spectral characterization process highly cumbersome and time-consuming and therefore severely limits the device throughput. To solve this long-standing issue, we develop a super-resolved nanophotonic snapshot hyper-spectral imaging (HSI) technique by taking advantage of wide-field PL imaging of quantum emitters embedded in a dispersive nanophotonic planar cavity, allowing simultaneous extractions of both spatial positions and emission wavelengths of a large number of single quantum emitters. 

\section{Principle of snapshot quantum HSI}

The concept of snapshot quantum HSI is presented in Fig.~\ref{fig:Fig1}. Spatial randomly distributed and spectral inhomogeneously broadened solid-state quantum emitters are placed in a 1D nanophotonic planar cavity, as shown in Fig.~\ref{fig:Fig1}(a). Different from localized micro-cavities in which the lights are highly confined in three dimensions, the nanophotonic planar cavity only provides light confinement in the Z-direction while the photons can freely propagate in the XY plane. The extended mode profile in the XY plane enables couplings between the cavity mode and multiple quantum emitters while the light confinement along the Z direction determines the dispersive feature of the cavity mode. Despite the Purcell factor provided by the planar cavity is very limited due to the rather large mode volume, such a cavity can significantly modify the radiation patterns of the embedded quantum emitters. Due to the dispersive couplings to the cavity modes, the emissions from solid-state quantum emitters (1-5) can only escape from the cavity at specific angles, determined by their emission wavelengths, thus exhibit characteristic far-field radiation patterns in the Fourier plane and unique profiles in the image plane, as presented in Fig.~\ref{fig:Fig1}(b). Based on the emission characteristics of single quantum emitters modified by the planar cavity mode, it is therefore possible to simultaneously extract the spatial positions ($\rm{x_{n}}, \rm{y_{n}}$) and emission wavelengths ($\rm{\lambda_{n}}$) of an ensemble of single quantum emitters by analyzing a snapshot PL image. With the extracted QD parameters, we could consequently realize a much stronger cavity QED effect by removing the original planar cavity and deterministically coupling QDs to 3D confined cavities with significantly reduced mode volumes and pre-designed cavity resonances, as shown in Fig.~\ref{fig:Fig1}(c).

\begin{figure*}
	\includegraphics[width=1\linewidth]{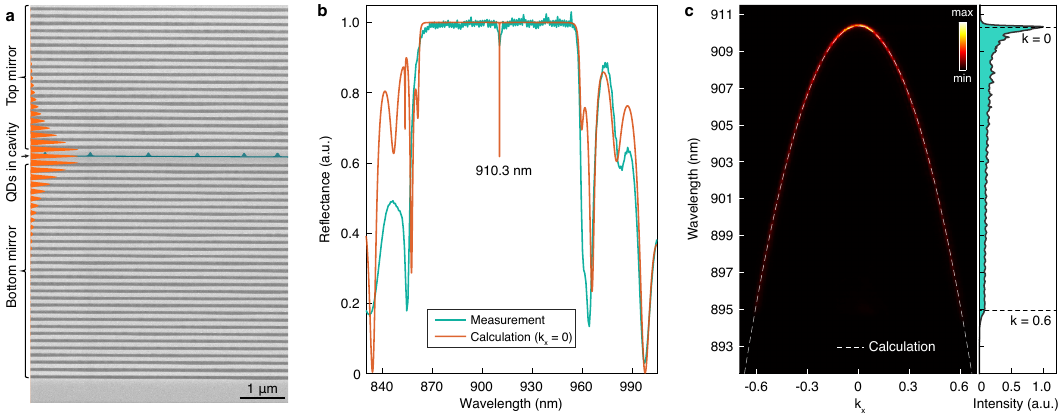}
	\caption{\textbf{Characterization of the dispersive semiconductor planar cavity.} (a) Scanning electron beam (SEM) image of the GaAs/AlGaAs planar cavity with a calculated mode profile (red) along Z direction superimposed. The InAs QDs are denoted by blue triangles. (b) Measurement (blue) and calculation (red) of the reflection spectrum in the normal (Z) direction. (c) Dispersion relation (cavity resonance as a function of the in-plane k-vector) of the planar cavity. Red solid line: angle-revolved PL spectra of the cavity mode. White dotted line: calculation with 1D TMM. Right side: integration of the angle-resolved cavity spectra in the momentum (k) space.}
	\label{fig:Fig2}
\end{figure*}

\section{Dispersion of a planar DBR cavity}

We now implement the concept of snapshot HSI to an III-V semiconductor system in which InAs QDs are embedded in a GaAs/AlGaAs planar cavity\cite{ramon2006emission,flagg2009resonantly,proux2015measuring,cogan2023deterministic}. Such coupled systems have been recognized as superior platforms for fabricating high-performance quantum light sources \cite{somaschi2016near,ding2016demand,bennett2016cavity,unsleber2016highly,he2017deterministic,liu2021dual,wei2022tailoring,gines2022high,coste2023high} and integrated quantum photonic circuits\cite{dusanowski2023chip}. As presented in Fig.~\ref{fig:Fig2}(a), an ensemble of InAs QDs grown by molecule beam epitaxy is embedded in a $\lambda$ GaAs layer sandwiched between a bottom and a top distributed Bragg reflector (DBR) consisting of 30 and 20 pairs of 1/4 $\lambda$ GaAs/$\rm Al_{0.9}Ga_{0.1}As$ layer, respectively. The resonant mode of the planar cavity is spectrally located in the middle of the DBR reflection band, as revealed from the reflection spectrum measured along the Z direction in Fig.~\ref{fig:Fig2}(b). To map the cavity resonance as a function of the in-plane k-vector (radiation direction), we perform angle-resolved spectroscopy~\cite{zhang2021momentum} on an ensemble of QDs under a high excitation power at which the QD ensemble serves as a broadband light source for probing the cavity mode dispersion, see the setups in Fig.~E1 of extended data. The radiation direction of the cavity mode is highly correlated to the cavity resonant wavelength and such a dispersion relation can be quantitatively reproduced by the 1D transfer-matrix method (TMM) (see methods), as shown in Fig.~\ref{fig:Fig2}(c). For comparison, the QDs in bulk material present no dispersion feature in both far-field patterns and image profiles, see Fig.~E2 of extended data.

\begin{figure*}
	\begin{center}
		\includegraphics[width=1\linewidth]{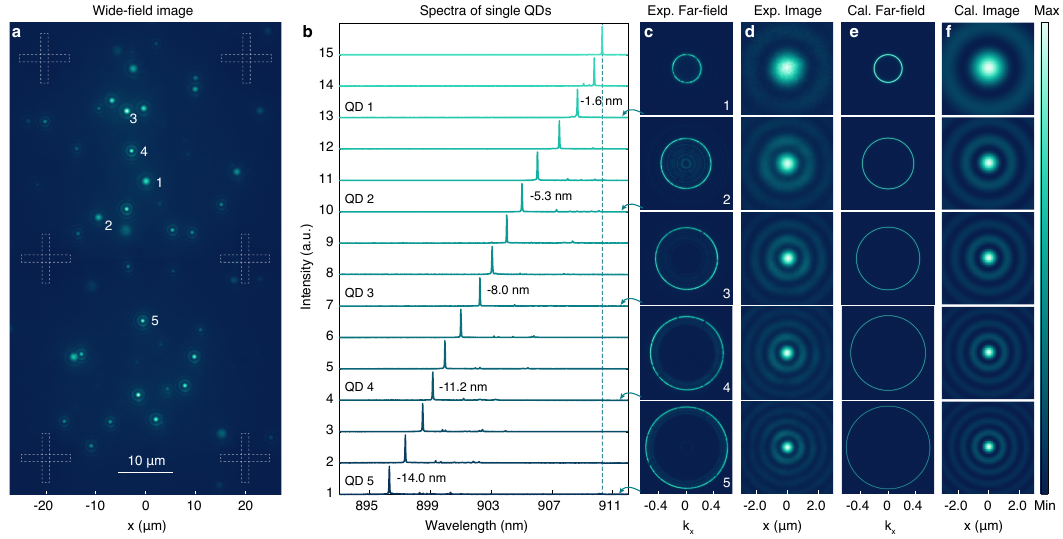}
		\caption{\textbf{Characterization of QDs in a planar cavity.}  (a) Wide-field image of an ensemble of QDs in a planar cavity. The crosses denote the metallic alignment marks used for the QD position extraction. (b) Confocal $\mu$PL spectra for 15 QDs in (a). The vertical dash line presents the cavity resonant wavelength measured along the Z direction. (c) Measured far-field patterns of 5 representative QDs (1-5). (d) Measured image profiles of 5 representative QDs (1-5). (e) Calculated far-field patterns of 5 representative QDs (1-5). (f) Calculated image profiles of 5 representative QDs (1-5).}
		\label{fig:Fig3}
	\end{center}
\end{figure*}

\section{Snapshot quantum HSI for QDs in a planar DBR cavity}

A typical wide-field PL image of the QD ensemble in the planar cavity is collected by a customized microscope and presented in Fig.~\ref{fig:Fig3}(a). Each individual QD presents itself as a bright spot at a particular position with a specific spot size in the wide-field PL image. The spatial positions of QDs respective to the pre-fabricated alignment marks can be extracted with an accuracy down to 20 nm ({see Fig.~E3 in the extended data}) by using the single molecule super-localization technique which is now widely employed for achieving spatial overlap between the QDs and the optical modes supported by the nanophotonic structures~\cite{liu2019solid,liu2021dual,wei2022tailoring}. Here, we focus on the relations between the emission wavelengths, far-field radiation patterns, and image profiles of different QDs. Fig.~\ref{fig:Fig3}(b) presents confocal 
$\mu$PL spectra of 15 individual QDs in the field-of-view, showing sharp exciton emission lines associated with single QDs. The QD spectra are displayed from bottom to top by the order of their emission wavelengths. The measured far-field radiation patterns and image profiles together with the calculations for 5 representative QDs (1-5) with different wavelengths are presented in Fig.~\ref{fig:Fig3}(c-f), respectively. The single QD emission lines exhibit ring-like far-field patterns due to the couplings to the planar cavity. As the QD emission wavelength moves towards the cavity resonance measured in the Z direction (dashed vertical line), the QD radiation angle (k-vector) decreases and forms an emission ring with a smaller diameter in the far-field, as presented in Fig.~\ref{fig:Fig3}(c,e). Such a behavior faithfully follows the dispersion characteristics of the planar cavity mode as shown in Fig.~\ref{fig:Fig2}(c). In contrast, the image profile of single QDs is in the form of an Airy beam consisting of a central bright spot surrounded by concentric rings with decreased intensities in Fig.~\ref{fig:Fig3}(d,f). Notably, those QDs with short emission wavelengths feature image sizes appreciably smaller than the standard diffraction limit set by the QD emission wavelength and the numerical aperture (NA) of the objetive~\cite{yang2013minimized} (see Fig.~E4 in the extended data), which provides a nanophotonic alternative of single-molecule super-resolution microscopy to the well-established technologies based on stimulated emission depletion (STED)\cite{vicidomini2018sted,liu2017amplified,jin2018nanoparticles,pu2022super}, structured illumination microscopy (SIM)\cite{heintzmann2009subdiffraction} and single-molecule localization microscopy (SMLM)\cite{lelek2021single,chi2020descriptor,pertsinidis2010subnanometre}. Due to the Fourier-transform relation between the far-field pattern and the image profile, the QD with a larger emission angle exhibits a smaller image size, which consequently correlates the emission wavelength to the image profile.

 \begin{center}
	\begin{figure*}
		\begin{center}
			\includegraphics[width=1\linewidth]{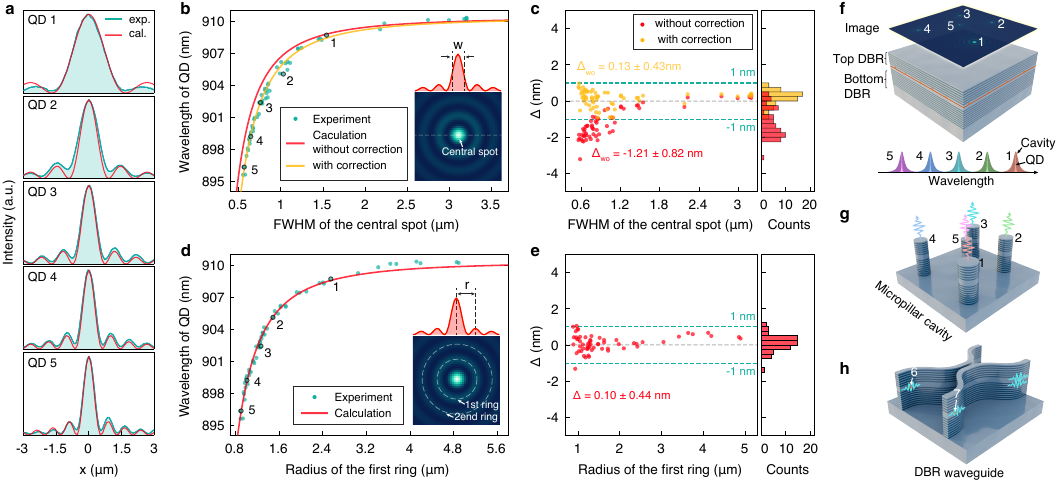}
			\caption{\textbf{Extraction of QD emission wavelengths from their image profiles.} (a) Crosscuts of the image profiles for QDs (1-5). (b) Relation between the QD emission wavelength and the FWHM (w) of the central spot. Green dots: experimental data. Red curve: calculation without any free parameter. Yellow curve: calculation with a background correction. (c) Deviations ($\Delta_{1}$) of the measured FWHM of the central spots from the theoretical curve for different QDs. Right: histogram of $\Delta_{1}$. (d) Relation between the QD emission and the radius of the first ring (r). Green dots: experimental data. Red curve: calculation without any free parameter. (e) Deviations ($\Delta_{2}$) of the measured radius of the first ring from the theoretical curve for different QDs. Right: histogram of $\Delta_{2}$. (f) Illustrations of HSI of QDs in a planar cavity for fabricating high-performance quantum light sources based on micropillars (g) and integrated photonic circuits based on DBR waveguides (h).}
			\label{fig:Fig4}
		\end{center}
	\end{figure*}
\end{center}

The crosscuts of the PL image profiles in Fig.~\ref{fig:Fig3}(d,~f) are presented in Fig.~\ref{fig:Fig4}(a) consisting of bright central peaks and weak side lobes. The FWHM of the central peak, w, and the radius of the first ring, r, (the distance between the central peak and the first side lobe) are closely correlated to the QD emission wavelength. We plot the emission wavelength as a function of w and compare the experimental results to the calculations in Fig.~\ref{fig:Fig4}(b). The red line is the calculation without any free parameter, which matches well with the trend of the experimental data but exhibits a slight offset. By performing a background subtraction in the model (yellow line), we achieve a perfect agreement between the experimental data and the calculation. The necessity of adding a correction in the model is probably due to the weak unwanted emission from other exciton states of QDs which leads to the over-estimation of w. The deviation ($\Delta_{1}=w_{\rm{exp}}-w_{\rm{cal}}$) of the measured w from the calculated values for QDs with different wavelengths are further presented in Fig.~\ref{fig:Fig4}(c), in which all the experimental data fall within the 1~nm range of the calculation with a mean value of 0.13~nm and a standard deviation of 0.43~nm when implementing the background subtraction. To avoid any post-correction in the modeling, we investigate the characteristic of the radius of the first ring, r, which is less sensitive to the unwanted background emission in the sample. Without any free parameter and background subtraction, we achieve an excellent agreement between the experiment and the calculation, as shown in Fig.~\ref{fig:Fig4}(d). The experimental data are well within the 1~nm range of the calculation, showing a deviation ($\Delta_{2}=r_{\rm{exp}}-r_{\rm{cal}}$) (Fig.~\ref{fig:Fig3}(e)) with a mean value of 0.1~nm and a standard deviation of 0.44~nm. Based on the relations obtained and verified in Fig.~\ref{fig:Fig4}(b,~d), QD emission wavelengths with an accuracy down to 0.1~nm can be reliably extracted from their image profiles. Therefore, we could simultaneously extract the emission wavelengths and spatial positions of a large number of QDs from a snapshot wide-field PL image. With the employment of snapshot HSI of QDs in a planar cavity (see Fig.~\ref{fig:Fig4}(f)), we could fabricate a large array of high-performance quantum light sources e.g., single-photon sources\cite{somaschi2016near,ding2016demand,bennett2016cavity,unsleber2016highly,he2017deterministic,liu2021dual,wei2022tailoring}, entangled photon pair sources\cite{gines2022high} and multi-photon entangled sources\cite{coste2023high}, via etching micropillar structures at the QD locations and targeting the cavity resonances to the QD emission wavelengths by controlling the sizes and geometries of the etched micropillars, as schematically shown in Fig.~\ref{fig:Fig4}(g) and experimentally verified in Fig.~E5. In addition, integrated quantum photonic circuits based on DBR waveguides, as shown in Fig.~\ref{fig:Fig4}(h), can be realized for pursuing on-chip Hong-Ou-Mandel interference between two QDs\cite{dusanowski2023chip}. Here we emphasize that the spectrometer is only used in Fig.~\ref{fig:Fig3}(b) to verify our method, it is not needed in the practical implementation of the HIS process.

\section{HSI based on opened planar DBR cavity for on-chip quantum circuit}

\begin{center}
	\begin{figure*}
		\begin{center}
			\includegraphics[width=1\linewidth]{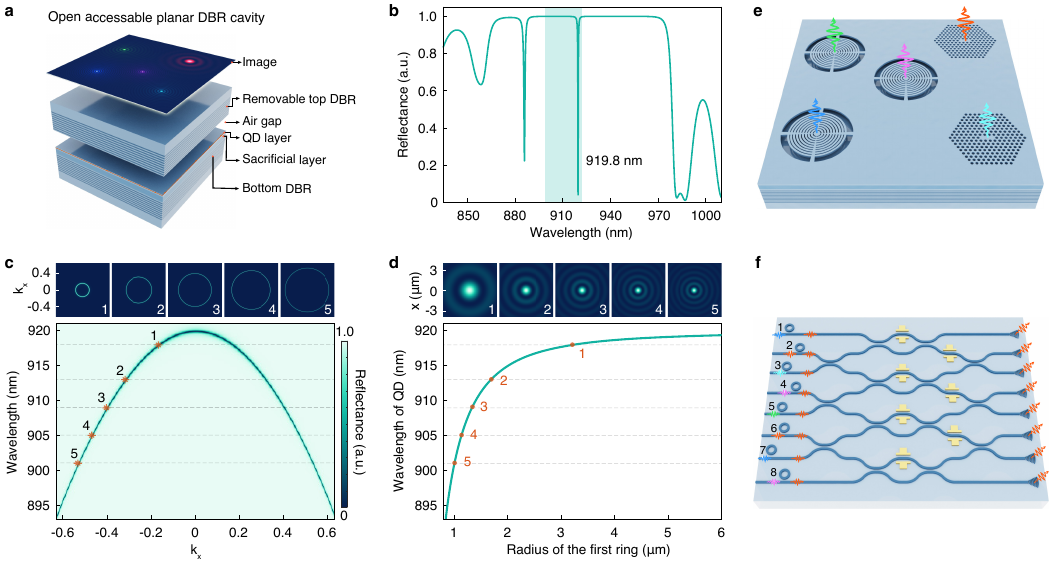}
			\caption{\textbf{Snapshot HSI of QDs in a planar open cavity for building thin-film QD devices with expanded functionalities and reduced footprints.}  (a) Schematics of the proposed planar open cavity for snapshot HSI. (b) Reflection spectrum in the normal direction for the planar open cavity. The blue area denotes the spectral range of the targeted QD emission. (c) Dispersion relation of the fundamental cavity mode used for HSI. Top insets: far-field radiation patterns for QDs with different wavelengths. (d) Relation between the emission wavelength and image profile for the QDs in the planar open cavity. Top insets: image profiles for QDs with different wavelengths. (e) Suspended circular Bragg resonators and photonic crystal cavities fabricated from the QDs in the planar open cavity by removing the AlGaAs sacrificial layer. (f) Large-scale integrated quantum photonic circuit consisting of quantum light sources, frequency converters, phase shifters, interferometers, and grating couplers fabricated from a planar open cavity by transferring thin-film GaAs layer with QDs on a low-index insulator substrate.}
			\label{fig:Fig5}
		\end{center}
	\end{figure*}
\end{center}

We further show that a proper modification of the cavity configuration could be exploited for building deterministically coupled thin-film QD devices with expanded functionalities and reduced footprints. The proposed planar open cavity is schematically shown in Fig.~\ref{fig:Fig5}(a). A 180~nm GaAs layer containing InAs QDs and a 450-nm-thick $\rm Al_{0.7}Ga_{0.3}As$ sacrificial layer are grown on a bottom DBR consisting of 20 pairs of 1/4 $\lambda$ GaAs/$\rm Al_{0.9}Ga_{0.1}As$ layer. A movable DBR mirror made of 20 pairs of 1/4 $\lambda$ GaAs/$\rm Al_{0.9}Ga_{0.1}As$ layer is placed on the top of the QD wafer with an air gap of 1000 nm, which could be precisely controlled by a piezo-nanopositioner\cite{tomm2021bright}. The normal reflection spectrum of the proposed planar open cavity is presented in Fig.~\ref{fig:Fig5}(b), exhibiting two cavity resonances in the stop band. We intentionally designed the fundamental cavity resonance to the proximity of the InAs QD emission wavelength range (895~nm - 920~nm) for HSI. The dispersion diagram of the targeted cavity mode is shown in Fig.~\ref{fig:Fig5}(c) in which the cavity resonance features a similar dispersion characteristic to that of the planar cavity in Fig.~\ref{fig:Fig2}(c). By using a microscope objective with an NA of 0.60 (corresponding to a divergence angle of 36.8 degrees), there is a one-to-one correspondence between the emission angle (k-vector) and the emission wavelength. Such a dispersion relation can be further mapped to the relation between the PL image profile and the QD emission wavelength in Fig.~\ref{fig:Fig5}(d) from which the QD emission wavelength can be directly extracted from the radius of the first ring in the PL image. With the spatial and spectral information of QDs obtained from the HSI, suspended membrane cavities with deterministically coupled QDs, such as circular Bragg resonators and high-Q photonic crystal cavities, can be fabricated for state-of-the-art quantum light sources and strong light-matter interactions (see Fig.~E6 of extended data), as schematically shown in Fig.~\ref{fig:Fig5}(e). To avoid the suspended membrane structures for large-scale integrations, more scalable devices such as microring quantum frequency convertors, phase shifters, interferometers and grating couplers, as shown in Fig.~\ref{fig:Fig5}(f), can also be pursued by transferring the thin GaAs layer containing QDs on a low-index insulator substrate\cite{castro2022expanding} (see Fig.~E7 of extended data).

\section{Conclusion}

To conclude, we’ve extended the HSI concept from classical optics to advanced quantum technologies by simultaneously extracting both spatial and spectral information of multiple solid-state quantum emitters from a snapshot wide-field PL image. We experimentally perform HIS of QDs in a planar cavity towards the realization of an extensive array of deterministically coupled QD-micropillar quantum light sources and integrated quantum photonic circuits based on DBR waveguides. To build integrated quantum photonic circuits with reduced footprints and expanded functionalities, such as frequency converters, phase shifters, interferometers and grating couplers, we propose and numerically demonstrate the feasibility of snapshot HSI of QDs in a planar open cavity. Our HSI technique can be straightforwardly adapted to other types of solid-state quantum emitters\cite{aharonovich2016solid}, e.g., NV centers, single molecules, and defects in GaN or 2D materials for high-throughput optical characterizations~\cite{sutula2023large}. More generally, it is also possible to implement snapshot HSI into biophysics where fast tracking and identification of single fluorescent particles are highly desirable yet technologically challenging. Our work provides an unprecedented tool to advance both classic optics and quantum photonics by enabling high-throughput deterministic light-matter interactions at the nanoscale.

\section{methods}
\noindent \textbf{Modeling:}
To calculate the far-field and image profile of the emission escaped from the planar DBR cavity, the parameters of the cavity are first extracted by fitting the measured reflectivity spectrum with a transform matrix method (TMM)~\cite{mackay2022transfer}, as shown in Fig.~\ref{fig:Fig2}(b). The cavity dispersion along the x-direction is then modeled by calculating the angle dependence reflectivity/transmission spectrum of the cavity (Fig.~\ref{fig:Fig5}(c)), which gives the relation between the QD emission wavelength and its momentum space distribution. For a specific wavelength, the calculated 1D far-field distribution is straightforwardly extended to a 2D far-field profile (XY plane)  by performing rotation transformation, considering the structure symmetry, as presented in Fig.~\ref{fig:Fig3}(e). The corresponding image profile is obtained by performing a Fourier transformation to the far-field profile within the angle defined by the NA of the objective. The FWHM of the central spot and the radius of the first airy ring are then extracted from the calculated image profile.

\noindent \textbf{Optical characterizations:}
The schematic of the setups for optical characterizations is presented in Extended Data Fig.~E1.

\vspace{0.8em}\noindent  \textbf{Data availability}

\noindent{The data that support the findings of this study are available within the paper and the Extended Data. Other relevant data are available from the corresponding authors on reasonable request.}

\vspace{0.8em}\noindent  \textbf{Acknowledgements}

\noindent{This research was supported by National Key Research and Development Program of China (2021YFA1400800); National Natural Science Foundation of China (62035017, 12304409), the Guangdong Special Support Program (2019JC05X397), and the National Super-Computer Center in Guangzhou.}

\vspace{0.8em}\noindent  \textbf{Author Contributions}

\noindent J.~L. conceived the project; S.~F.~L., J.~L., and H.~Q.~L. designed the epitaxial structure and the devices; H.~Q.~L. and H.~Q.~N. grew the quantum dot wafers; S.~F.~L. and G.~X.~Q. developed the theory model and calculated the image and far-field profile; S.~F.~L. and X.~S.~L. fabricated the devices; S.~F.~L. built the setup and performed the optical measurements with inputs from X.~S.~L., J.~T.~M., and L.~N.; S.~F.~L. and J.~L. analyzed the data; J.~L. and S.~F.~L. prepared the manuscript with inputs from all authors; J.~L., Z.~C.~N., C.-W.~Q. and X.~H.~W. supervised the project.

\vspace{0.8em}\noindent  \textbf{Conflict of Interest}

\noindent The authors declare no conflict of interest.

\end{document}